\documentstyle[12pt]{article}
\begin{document}
\title{STRUCTURE AND EVOLUTION OF  SELF--GRAVITATING OBJECTS AND THE ORTHOGONAL SPLITTING OF THE RIEMANN TENSOR}
\author{L. Herrera$^1$\thanks{Postal
address: Apartado 80793, Caracas 1080A, Venezuela.} \thanks{e-mail:
laherrera@cantv.net.ve}, 
 J. Ospino$^2$\thanks{e-mail:
jhozcra@usal.es}, A. Di Prisco$^1$\thanks{e-mail:
adiprisc@fisica.ciens.ucv.ve}
E. Fuenmayor$^1$ \thanks{e-mail: efuenma@fisica.ciens.ucv.ve} \\and  O. Troconis$^1$\thanks{e-mail:otroconis@fisica.ciens.ucv.ve}
\\
{\small $^1$Escuela de F\'{\i}sica, Facultad de Ciencias,}\\
{\small Universidad Central de Venezuela, Caracas, Venezuela.} \\
{\small $^2$Area de F\'{\i}sica Te\'orica, Facultad de Ciencias,}\\ 
{\small Universidad de Salamanca, 37008 Salamanca, Espa\~na.}\\
}
\maketitle
\newpage
\begin{abstract}
The full set of equations governing the structure and the  evolution of self--gravitating spherically symmetric dissipative fluids with anisotropic stresses, is written down in terms of  five scalar quantities obtained from the orthogonal splitting of the Riemann tensor,  in the context of general relativity.  It is shown that these scalars are directly related to fundamental properties of the fluid distribution, such as: energy density, energy density inhomogeneity, local anisotropy of pressure, dissipative flux and the active gravitational mass. It is also shown that in the static case, all possible solutions to Einstein equations may be expressed explicitly through these scalars. Some solutions are  exhibited to illustrate this point.
\end{abstract}
\pagebreak
\section{INTRODUCTION}
 The gravitational collapse of massive stars, and  its resulting remanent (neutron star or black hole) represent one of the few observable scenarios  where general relativity is expected to
play a relevant role. Therefore,   a detailed description of gravitational collapse and the modelling of the structure of compact objects under a variety of conditions, remain among the most  interesting problems that general relativity has to deal with. This fact explains  the great attraction that these problems  exert on the community of the relativists.  Starting with   the seminal papers by Oppenheimer
and Snyder \cite{Opp} and Oppeheimer and Volkoff \cite{OppII}, a long list  of works  has been presented trying to provide  models of evolving gravitating spheres and compact objects (just as a sample, and without  the pretension of being exhaustive, see \cite{May}--\cite{mis23} and references therein).

Motivated by the above arguments, we present  in this work a  study on self--gravitating relativistic fluids  in terms of a set of scalars obtained from the orthogonal splitting of the Riemann tensor. As we shall see these scalars have a distinct physical meaning and appear to be particularly well suited for the description of self--gravitating fluids.

We shall assume our system to be spherically symmetric. This is a common assumption in the study of self--gravitating compact objects, since  deviations from spherical
symmetry are likely to be incidental rather than  basic features of the process involved (see however the discussion in \cite{nsp1}--\cite{nsp3}). 

For sake of generality we shall further assume  the fluid to be locally anisotropic (principal stresses unequal) and dissipative.

The assumption of local anisotropy of pressure, which seems to be very sensible to describe matter distribution under a variety of circumstances, has been proved
to be very useful in the study of relativistic compact objects and related problems
(see
\cite{PR} for a comprehensive review until 1997, and \cite{catoen}--\cite{mos} and references therein, for more recent developments). 

On the other hand, it is already an established fact, that gravitational collapse is a
highly dissipative process (see \cite{Hs}, \cite{Mitra}, \cite{Hetal} and references
therein). This dissipation is required to account for the very large
(negative) binding energy of the resulting compact object (of the order
of $-10^{53} erg$).

In fact, it seems that the only plausible
mechanism to carry away the bulk of the binding energy of the collapsing
star, leading to a neutron star or black hole, is neutrino emission
\cite{1}. 

As usual, we shall describe dissipation in  two limiting cases.  The first case (diffusion) applies whenever  the mean free path of
particles responsible for the propagation of energy  is  very small as compared with the typical
length of the object.
For example, for a main sequence star as the sun, the mean free path of
photons at the center, is of the order of $2$ cm. Also, the
mean free path of trapped neutrinos in compact cores of densities
about $10^{12}$ g. cm$^{-3}$ becomes smaller than the size of the stellar
core \cite{3,4}. Furthermore, the observational data collected from supernova 1987A
indicates that the regime of radiation transport prevailing during the
emission process, is closer to the diffusion approximation than to the free
streaming limit \cite{5}.
In this case, it is assumed that the energy flux of
radiation, as that of
thermal conduction, is proportional to the gradient of temperature.

The second case (free streaming), applies when the mean free path of particles
transporting energy is larger than (or equal to) the typical length of the object. Since this condition may hold for a large number of stellar scenarios, it is advisable to include simultaneously both limiting  cases of radiative transport, diffusion and free streaming, allowing for describing a wide range of
situations.

In the next  section we shall introduce the relevant physical variables and deploy the equations for describing a dissipative self--gravitating locally anisotropic fluid.

As mentioned  before, a fundamental role in our study will be played by a set of scalars derived from the orthogonal splitting of the Riemann tensor. Such a splitting and the ensuing variables, are presented in  Section III. In Section IV the physical meaning of the above mentioned scalars is discussed, and in Section V a general method to obtain all static anisotropic solutions in terms of those scalars, is presented.

Finally the results are discussed in the  last section.

\section{THE GENERAL FORMALISM}
 
In this section we shall present the physical variables and the relevant equations for describing a dissipative self--gravitating locally anisotropic fluid. Here we shall closely follow the program outlined in \cite{Hetal}, thus we refer the reader to that article for more details.
\subsection{Einstein equations}

We consider spherically symmetric distributions of collapsing
fluid, which for the sake of completeness we assume to be locally anisotropic,
undergoing dissipation in the form of heat flow (diffusion limit)  and/or free streaming  radiation
 (free streaming limit), bounded by a
spherical surface $\Sigma$.

\noindent
The line element is given in Schwarzschild--like (non--comoving)  coordinates by

\begin{equation}
ds^2=e^{\nu} dt^2 - e^{\lambda} dr^2 -
r^2 \left( d\theta^2 + \sin^2\theta d\phi^2 \right),
\label{metric}
\end{equation}

\noindent
where $\nu(t,r)$ and $\lambda(t,r)$ are functions of their arguments. We
number the coordinates: $x^0=t; \, x^1=r; \, x^2=\theta; \, x^3=\phi$.

\noindent
The metric (\ref{metric}) has to satisfy Einstein field equations

\begin{equation}
G^\nu_\mu=\kappa T^\nu_\mu,
\label{Efeq}
\end{equation}

\noindent
with $\kappa=8\pi$ and which in our case read \cite{May}:
\begin{equation}
-\kappa T^0_0 =-{1 \over r^2} +e^{-\lambda} ({1 \over r^2}
-{\lambda ^{\prime} \over r}),
\label{1}
\end{equation}
\begin{equation}
-\kappa T^1_1=-{1 \over r^2} +e^{-\lambda} ({1 \over r^2} +{{\nu
^{\prime}} \over r}),
\label{2}
\end{equation}
$$-\kappa T^2_2=-\kappa T^3_3=-\frac{e^{-\nu}}{4} (2 \ddot
\lambda+\dot \lambda  (\dot \lambda -\dot \nu ))$$
\begin{equation}
+\frac{e^{-\lambda}}{4}(2 \nu ^{\prime \prime} + {\nu
^{\prime}}^2-\lambda ^{\prime} \nu ^{\prime} + 2 \frac{\nu
^{\prime}-\lambda ^{\prime}}{r}),
\label{3'}
\end{equation}
\begin{equation}
-\kappa T_{10} =-\frac{\dot \lambda}{r},
\label{4}
\end{equation}
\noindent
where dots and primes stand for partial differentiation with respect
to $t$ and $r$,
respectively.
\noindent
In order to give physical significance to the $T^{\mu}_{\nu}$ components
we apply the Bondi approach \cite{May}.

\noindent
Thus, following Bondi, let us introduce purely locally Minkowski
coordinates ($\tau, x, y, z$)

$$d\tau=e^{\nu/2}dt\,;\qquad\,dx=e^{\lambda/2}dr\,;\qquad\,
dy=rd\theta\,;\qquad\, dz=r\sin \theta d\phi.$$

\noindent
Then, denoting the Minkowski components of the energy tensor by a bar,
we have

$$\bar T^0_0=T^0_0\,;\qquad\,
\bar T^1_1=T^1_1\,;\qquad\,\bar T^2_2=T^2_2\,;\qquad\,
\bar T^3_3=T^3_3\,;\qquad\,\bar T_{01}=e^{-(\nu+\lambda)/2}T_{01}.$$

\noindent
Next, we suppose that when viewed by an observer moving relative to these
coordinates with proper velocity $\omega$ in the radial direction, the physical
content  of space consists of an anisotropic fluid of energy density $\rho$,
radial pressure $P_r$, tangential pressure $P_\bot$,  radial heat flux
$q$ and unpolarized radiation of energy density $\epsilon$
traveling in the radial direction. Thus, when viewed by this (comoving with the fluid)
observer the covariant energy--momentum tensor in
Minkowski coordinates is

\[ \left(\begin{array}{cccc}
\rho + \epsilon    &  - q -\epsilon  &   0     &   0    \\
- q - \epsilon &  P_r + \epsilon    &   0     &   0    \\
0       &   0       & P_\bot  &   0    \\
0       &   0       &   0     &   P_\bot
\end{array} \right). \]

\noindent
Then a Lorentz transformation gives
\begin{equation}
T^0_0= \bar T^0_0 =\frac{\rho+P_r \omega ^2}{1-\omega
^2}+\frac{2\omega
q}{1-\omega^2}+\frac{\epsilon(1+\omega)}{1-\omega},
\label{8'}
\end{equation}
\begin{equation}
T^1_1=\bar T^1_1 =-\frac{P_r+\rho \omega ^2}{1-\omega
^2}-\frac{2\omega
q}{1-\omega^2}-\frac{\epsilon(1+\omega)}{1-\omega},
\label{9}
\end{equation}
\begin{equation}
T^2_2=T^3_3=\bar T^2_2 =\bar T^3_3 =-P_{\bot},
\label{10}
\end{equation}
\begin{equation}
T_{01}=e^{\frac{\nu+\lambda}{2}}\bar
T_{01}=-\frac{(\rho+P_r)\omega e^{\frac{\nu+\lambda}{2}}}{1-\omega
^2}-\frac{q e^{\frac{\lambda+\nu}{2}}}{1-\omega ^2}(1+\omega
^2)-\frac{e^{\frac{\lambda +\nu}{2}}
\epsilon(1+\omega)}{1-\omega}.
\label{11a}
\end{equation}

\noindent
Note that the coordinate velocity in the ($t,r,\theta,\phi$) system, $dr/dt$,
is related to $\omega$ by

\begin{equation}
\omega=\frac{dr}{dt}\,e^{(\lambda-\nu)/2}.
\label{omega}
\end{equation}

\noindent
Feeding back (\ref{8'}--\ref{11a}) into (\ref{1}--\ref{4}), we get
the field equations in  the form

\begin{equation}
\frac{\rho + P_r \omega^2 }{1 - \omega^2} +\frac{2\omega
q}{1-\omega^2}+\frac{\epsilon(1+\omega)}{1-\omega}=-\frac{1}{\kappa}\Biggl\{-\frac{1}{r^2}+e^{-\lambda}
\left(\frac{1}{r^2}-\frac{\lambda'}{r} \right)\Biggr\},
\label{fieq00}
\end{equation}

\begin{equation}
\frac{ P_r + \rho \omega^2}{1 - \omega^2} +
\frac{2\omega
q}{1-\omega^2}+\frac{\epsilon(1+\omega)}{1-\omega}=-\frac{1}{\kappa}\Biggl\{\frac{1}{r^2} - e^{-\lambda}
\left(\frac{1}{r^2}+\frac{\nu'}{r}\right)\Biggr\},
\label{fieq11}
\end{equation}

\begin{eqnarray}
P_\bot = -\frac{1}{\kappa}\Biggl\{\frac{e^{-\nu}}{4}\left(2\ddot\lambda+
\dot\lambda(\dot\lambda-\dot\nu)\right) \nonumber \\
 - \frac{e^{-\lambda}}{4}
\left(2\nu''+\nu'^2 -
\lambda'\nu' + 2\frac{\nu' - \lambda'}{r}\right)\Biggr\},
\label{fieq2233}
\end{eqnarray}

\begin{equation}
\frac{(\rho + P_r) \omega e^{\frac{\lambda +\nu}{2}}}{1 - \omega^2} +\frac{q e^{\frac{\lambda+\nu}{2}}}{1-\omega ^2}(1+\omega
^2)+\frac{e^{\frac{\lambda +\nu}{2}}
\epsilon(1+\omega)}{1-\omega}=-\frac{\dot\lambda}{\kappa r}.
\label{fieq01}
\end{equation}

Next, the four--velocity  vector is defined as
\begin{equation}
u^{\alpha}=(\frac{e^{-\frac{\nu}{2}}}{(1-\omega^2)^{1/2}},\frac{\omega
e^{ -\frac{\lambda}{2}}}{(1-\omega^2)^{1/2}},0,0),
\end{equation}
 from which we can calculate the four acceleration, $a^\alpha=u^\alpha_{;\beta}u^\beta$ 
\begin{equation}
\omega a_1=-a_0 e^{\frac{\lambda-\nu}{2}}=-\frac{\omega}{1-\omega
^2}\left [(\frac{\omega \omega ^{\prime}}{1-\omega^2}+\frac{\nu
^{\prime}}{2})+e^{\frac{\lambda-\nu}{2}}(\frac{\omega \dot
\lambda}{2}+\frac{\dot \omega}{1-\omega^2})\right],
\label{15a}
\end{equation}
the shear tensor $\sigma_{\mu \nu}$ and the expansion $\Theta$

\begin{equation}
\sigma_{\mu\nu}=u_{\mu;\nu}+u_{\nu;\mu}-u_{\mu}a_{\nu}-u_{\nu}a_{\mu}-
\frac{2}{3}{\Theta}h_{\mu\nu},
\label{shear}
\end{equation}

\noindent
with

\begin{equation}
h_{\mu\nu}=g_{\mu\nu}-u{_\mu}u_{\nu}\,\qquad\,
\Theta=u^{\mu}_{;\mu},
\label{P,Theta,a}
\end{equation}

\begin{equation}
\Theta=\frac{e^{-\nu/2}}{2\left(1-\omega^2\right)^{1/2}}
\left(\dot\lambda + \frac{2\omega\dot\omega}{1-\omega^2}\right) + 
\frac{e^{-\lambda/2}}{2\left(1-\omega^2\right)^{1/2}}
\left(\omega\nu' + \frac{2\omega'}{1-\omega^2}  + 
\frac{4\omega}{r}\right),
\label{Theta}
\end{equation} 

\begin{equation}
\sigma_{11}= -\frac{2}{3\left(1-\omega^2\right)^{3/2}}
\left[e^{\lambda}e^{-\nu/2}\left(\dot\lambda + 
\frac{2\omega\dot\omega}{1-\omega^2}\right) + e^{\lambda/2} 
\left(\omega\nu' + \frac{2\omega'}{1-\omega^2} 
- \frac{2\omega}{r}\right)  \right],
\label{shear11}
\end{equation}

\begin{equation}
\sigma_{22}=-\frac{e^{-\lambda} r^2 \left(1-\omega^2\right)}{2}
\sigma_{11},
\label{shear22}
\end{equation}

\begin{equation}
\sigma_{33}=-\frac{e^{-\lambda} r^2 \left(1-\omega^2\right)}{2} 
\sin^2{\theta} \sigma_{11},
\label{shear33}
\end{equation}

\begin{equation}
\sigma_{00}=\omega^2 e^{-\lambda} e^\nu \sigma_{11},
\label{shear00}
\end{equation}
\begin{equation}
\sigma_{01}=-\omega e^{\left(\nu-\lambda\right)/2} \sigma_{11}.
\label{shear01}
\end{equation}
 We may write the shear tensor  also as 
\begin{equation}
 \sigma _{\alpha \beta}=\frac{1}{2}\sigma (s_\alpha
s_\beta+\frac{1}{3}h_{\alpha_\beta}),
\label{52}
\end{equation}

with

\begin{equation}
\sigma=-\frac{1}{(1-\omega^2)^{1/2}}\left[e^{-\frac{\nu}{2}}(\dot
\lambda+\frac{2\omega \dot \omega}{1-\omega
^2})+e^{-\frac{\lambda}{2}}(\omega \nu ^{\prime}+\frac{2\omega
^{\prime}}{1-\omega ^2}-\frac{2\omega}{r})\right],
\label{sigma}
\end{equation}

and  $s^\mu$  being defined by
\begin{equation}
s^{\mu}=(\frac{\omega
e^{-\frac{\nu}{2}}}{(1-\omega^2)^{1/2}},
\frac{e^{-\frac{\lambda}{2}}}{(1-\omega^2)^{1/2}},0,0)\label{ese},
\end{equation}

with the  properties
$s^{\mu}u_{\mu}=0$,
$s^{\mu}s_{\mu}=-1$.

It will be convenient to write the energy--momentum tensor  (\ref{8'})-(\ref{11a}) as:
\begin{equation}
T^{\mu}_{\nu}=\tilde \rho u^{\mu}u_{\nu}- \hat P
h^{\mu}_{\nu}+\Pi ^{\mu}_{\nu} +\tilde q(s^\mu u_\nu+s_\nu u^\mu),
\label{24'}
\end{equation}
with $$\Pi
^{\mu}_{\nu}=\Pi(s^{\mu}s_{\nu}+\frac{1}{3}h^{\mu}_{\nu}),$$
$$\tilde q^{\mu}=\tilde qs^{\mu},$$
$$\hat P=\frac{\tilde P_{r}+2P_{\bot}}{3},$$

$$\tilde \rho= \rho+\epsilon,$$
$$\tilde P_{r}=P_r+\epsilon,$$
$$\tilde q= q+\epsilon,$$
$$\Pi=\tilde P_{r}-P_{\bot}.$$

\noindent
For the exterior of the fluid distribution, the spacetime is that of Vaidya,
given by

\begin{equation}
ds^2= \left(1-\frac{2M(u)}{\cal R}\right) du^2 + 2dud{\cal R} -
{\cal R}^2 \left(d\theta^2 + \sin^2\theta d\phi^2 \right),
\label{Vaidya}
\end{equation}

\noindent
where $u$ is a coordinate related to the retarded time, such that
$u=constant$ is (asymptotically) a
null cone open to the future and $\cal R$ is a null coordinate ($g_{
\cal R\cal R}=0$). 
\noindent

The two coordinate systems ($t,r,\theta,\phi$) and ($u,
\cal R,\theta,\phi$) are
related at the boundary surface  by

\begin{equation}
u=t-r-2M \ln \left(\frac{r}{2M}-1\right),
\label{u}
\end{equation}

\begin{equation}
{\cal R}=r.
\label{radial}
\end{equation}

\noindent
In order to match smoothly the two metrics above on the boundary surface
$r=r_\Sigma(t)$, we  require the continuity of the first and the second fundamental
forms across that surface, yielding (see \cite{HJR} for details)
\begin{equation}
e^{\nu_\Sigma}=1-\frac{2M}{{\cal R}_\Sigma},
\label{enusigma}
\end{equation}
\begin{equation}
e^{-\lambda_\Sigma}=1-\frac{2M}{{\cal R}_\Sigma},
\label{elambdasigma}
\end{equation}
\begin{equation}
\left[P_r\right]_\Sigma=\left[q \right]_\Sigma ,
\label{PQ}
\end{equation}
where, from now on, subscript $\Sigma$ indicates that the quantity is
evaluated on the boundary surface $\Sigma$, and (\ref{PQ}) expresses the discontinuity of the radial pressure in the presence
of heat flow, which is a well known result \cite{Sa}.

Eqs. (\ref{enusigma}), (\ref{elambdasigma}), and (\ref{PQ}) are the necessary and
sufficient conditions for a smooth matching of the two metrics (\ref{metric})
and (\ref{Vaidya}) on $\Sigma$.

\subsection{The Riemann and the Weyl tensor}
 We know that the
Riemann tensor may be expressed through the Weyl tensor
$C^{\rho}_{\alpha
\beta
\mu}$, the  Ricci tensor $R_{\alpha\beta}$ and the scalar curvature $R$,
as:
$$
R^{\rho}_{\alpha \beta \mu}=C^\rho_{\alpha \beta \mu}+ \frac{1}{2}
R^\rho_{\beta}g_{\alpha \mu}-\frac{1}{2}R_{\alpha \beta}\delta
^\rho_{\mu}+\frac{1}{2}R_{\alpha \mu}\delta^\rho_\beta$$
\begin{equation}
-\frac{1}{2}R^\rho_\mu g_{\alpha
\beta}-\frac{1}{6}R(\delta^\rho_\beta g_{\alpha \mu}-g_{\alpha
\beta}\delta^\rho_\mu).
\label{34}
\end{equation}

In the spherically symmetric case, the magnetic part of the Weyl tensor vanishes and we can express the Weyl tensor in terms of its electric part ($E_{\alpha \beta}=C_{\alpha \gamma \beta
\delta}u^{\gamma}u^{\delta}$) as
\begin{equation}
C_{\mu \nu \kappa \lambda}=(g_{\mu\nu \alpha \beta}g_{\kappa \lambda \gamma
\delta}-\eta_{\mu\nu \alpha \beta}\eta_{\kappa \lambda \gamma
\delta})u^\alpha u^\gamma E^{\beta \delta},
\label{40}
\end{equation}
with $g_{\mu\nu \alpha \beta}=g_{\mu \alpha}g_{\nu \beta}-g_{\mu
\beta}g_{\nu \alpha}$,   and $\eta_{\mu\nu \alpha \beta}$ denoting the Levi--Civita tensor.
Observe that $E_{\alpha \beta}$ may also be writen as 
\begin{equation}
E_{\alpha \beta}=E (s_\alpha s_\beta+\frac{1}{3}h_{\alpha \beta}),
\label{52bisx}
\end{equation}
with
\begin{eqnarray}
E&=&\frac{e^{-\nu}}{4} \left[ \ddot
\lambda+\frac{\dot \lambda (\dot \lambda -\dot \nu )}{2}\right] \nonumber\\
&-&\frac{e^{-\lambda}}{4}\left[ \nu ^{\prime \prime} + \frac{{\nu
^{\prime}}^2-\lambda ^{\prime} \nu ^{\prime}}{2} -  \frac{\nu
^{\prime}-\lambda ^{\prime}}{r}+\frac{2(1-e^{\lambda})}{r^2}\right],
\label{defE}
\end{eqnarray}
satisfying the following properties:
 \begin{eqnarray}
 E^\alpha_{\,\,\alpha}=0,\quad E_{\alpha\gamma}=
 E_{(\alpha\gamma)},\quad E_{\alpha\gamma}u^\gamma=0.
  \label{propE}
 \end{eqnarray} 

\subsection{The mass function and the Tolman mass}
Here we shall introduce the two most commonly used definitions for the mass of a sphere interior to the surface $\Sigma$, as well as some interesting relationships between them and the Weyl tensor. These will be used later to provide physical meaning to the five scalars quantities which will be derived from the orthogonal splitting of the Riemann tensor.
\subsubsection{The mass function}
For the line element (\ref{metric}) the mass function $m$ is defined by:
\begin{equation}
R^3_{232}=1-e^{-\lambda}=\frac{2m}{r}.
\label{rieman}
\end{equation}

Then, using  (\ref{Efeq}), (\ref{34}) and (\ref{52bisx})  we may write
 \begin{equation}
\frac{3m}{r^3}=\frac{\kappa}{2} \tilde \rho +\frac{\kappa}{2} (P_\bot -\tilde P_r)+E.
\label{66}                   
\end{equation}
Observe that from (\ref{1}) and (\ref{rieman}) the mass function  may also be writen  as
\begin{equation}
m = \frac{\kappa}{2} \int^{r}_{0} r^2 T^0_0 dr.
\label{m}
\end{equation}

Another interesting relationship for the mass function may be obtained as follows. From  field equations (\ref{fieq00})--(\ref{fieq2233}), and (\ref{34}), (\ref{52bisx}) and (\ref{m}) we get

\begin{equation}
m = \frac{\kappa}{6} r^3 \left(T^0_0 + T^1_1 - T^2_2\right) + \frac{r^3 E}{3}.
\label{mW}
\end{equation}

\noindent
Next, differentiating (\ref{mW}) with respect to $r$ and using (\ref{m}) 
it follows

\begin{equation}
\left(\frac{r^3 E}{3}\right)' = - \frac{\kappa}{6} r^3 \left(T^0_0\right)' + 
\frac{\kappa}{6} \left[r^3 \left(T^2_2 - T^1_1\right)\right]' ,
\label{Wpri}
\end{equation}

\noindent
and integrating

\begin{equation}
 E= - \frac{\kappa}{2r^3} \int^r_0{r^3 \left(T^0_0\right)' dr} + 
\frac{\kappa}{2} \left(T^2_2 - T^1_1\right).
\label{Wint}
\end{equation}

\noindent
Finally, inserting (\ref{Wint}) into (\ref{mW}) we obtain

\begin{equation}
m(r,t) = \frac{\kappa}{6} r^3 T^0_0 - 
\frac{\kappa}{6} \int^r_0{r^3 \left(T^0_0\right)'dr}.
\label{mT00}
\end{equation}

\noindent Now, there are three  specific situations when $T^0_0=\tilde \rho$ and $T^2_2 - T^1_1=\Pi$, namely:
\begin{itemize}
\item In the static regime, i.e. when $\omega$ as well as all time derivatives vanish.
\item In the quasi--static regime, where  (see \cite{HJR}) \begin{equation}
\omega^2 \approx \dot \omega \approx \ddot \nu \approx \ddot\lambda \approx \dot\nu \dot\lambda \approx
\dot\lambda^2 \approx 0
\label{lp0}
\end{equation}
\item Immediately after the system departs from equilibrium,  i.e. on a time scale of  the order of  (or smaller than) the hydrostatic  time, in which  case  $\omega\approx \dot \lambda \approx  \dot \nu\approx 0; \dot \omega \neq 0$.
\end{itemize}

Thus in the three cases above, (\ref{Wint}) and (\ref{mT00}) become
\begin{equation}
 E= - \frac{\kappa}{2r^3} \int^r_0{r^3 \left(\tilde \rho\right)' dr} + 
\frac{\kappa}{2} \Pi,
\label{Wintbis}
\end{equation}

\begin{equation}
m(r,t) = \frac{\kappa}{6} r^3 \tilde \rho - 
\frac{\kappa}{6} \int^r_0{r^3 \left(\tilde \rho \right)'dr}.
\label{mT00bis}
\end{equation}

The first of these equations relates the Weyl tensor to two fundamental physical properties of the fluid distribution, namely: density inhomogeneity and local anisotropy of pressure.
The second one expresses the mass function in  terms of its value in the case of a homogeneous energy density distribution, plus the change induced by density inhomogeneity.
\subsubsection{Tolman mass}
An alternative definition to  describe the energy content  of a  fluid sphere  was proposed by Tolman many years ago.
\noindent
The Tolman mass for a spherically symmetric distribution 
of matter is given by (eq.(24) in \cite{To})

\begin{eqnarray}
m_T = & &  \frac{\kappa}{2} \int^{r_\Sigma}_{0}{r^2 e^{(\nu+\lambda)/2} 
\left(T^0_0 - T^1_1 - 2 T^2_2\right) dr}\nonumber \\ 
& + & \frac{1}{2} \int^{r_\Sigma}_{0}{r^2 e^{(\nu+\lambda)/2} 
\frac{\partial}{\partial t} 
\left(\frac{\partial L}{\partial \left[\partial 
\left(g^{\alpha \beta} \sqrt{-g}\right) / \partial t\right]}\right) g^{\alpha \beta}dr},
\label{Tol}
\end{eqnarray}

\noindent
where $L$ denotes the usual gravitational Lagrangian density 
(eq.(10) in \cite{To}). Although Tolman's formula was introduced 
as a measure of the total energy of the system, with no commitment 
to its localization, we shall define the mass within a sphere of 
radius $r$, completely inside $\Sigma$, as 

\begin{eqnarray}
m_T = & &  \frac{\kappa}{2} \int^{r}_{0}{r^2 e^{(\nu+\lambda)/2} 
\left(T^0_0 - T^1_1 - 2 T^2_2\right) dr}\nonumber \\ 
& + & \frac{1}{2} \int^{r}_{0}{r^2 e^{(\nu+\lambda)/2} 
\frac{\partial}{\partial t} 
\left(\frac{\partial L}{\partial \left[\partial 
\left(g^{\alpha \beta} \sqrt{-g}\right) / \partial t\right]}\right) g^{\alpha \beta}dr}.
\label{Tolin}
\end{eqnarray}

\noindent
This extension of the global concept of energy to a local level 
\cite{Coo} is suggested by the conspicuous role played by 
$m_T$ as the ``active gravitational mass'', which will be 
exhibited below.

Now, it can be shown after some lengthy calculations (see \cite{inh} for details) that 
 
\begin{equation}
m_T  = e^{(\nu + \lambda)/2} 
\left[m(r,t) - \frac{\kappa}{2} r^3 T^1_1\right].
\label{I+II}
\end{equation}

\noindent
Replacing $T^1_1$ by (\ref{2}) and $m$ by (\ref{rieman}), 
one  also finds

\begin{equation}
m_T = e^{(\nu - \lambda)/2} \, \nu' \, \frac{r^2}{2}.
\label{mT}
\end{equation}

\noindent
This last equation brings out the physical meaning of $m_T$ as the 
active gravitational mass. Indeed, as it follows from (\ref{15a}), the gravitational acceleration ($a=-s^\nu a_\nu$) of a test particle, 
instantaneously at rest in a static gravitational field,  is given by (see also \cite{Gro})

\begin{equation}
a = \frac{e^{- \lambda/2} \, \nu'}{2} =  \frac{e^{-\nu/2}m_T}{r^2} .
\label{a}
\end{equation}

Another expression for $m_T$, 
which appears to be more suitable for the discussion in Sec.IV may be obtained as follows. 
Taking the $r$-derivative of (\ref{mT}) (see \cite{inh} for details, but notice some minor misprints in eqs.(31) and (38) in that reference as well as slight changes in notation) and using (\ref{mW}), 
(\ref{I+II}), and field equations, we obtain

\begin{eqnarray}
r m'_T - 3 m_T & = & e^{(\nu + \lambda)/2} r^3 \left[
\frac{\kappa}{2} \left(T^1_1 - T^2_2\right) - E\right] \nonumber \\
& + &
\frac{e^{(\lambda - \nu)/2} r^3}{2} \left(\ddot\lambda + 
\frac{\dot\lambda^2}{2} - \frac{\dot\lambda \dot\nu}{2}\right),
\label{pre}
\end{eqnarray}
which can be formally integrated to give

\begin{eqnarray}
m_T & = & (m_T)_\Sigma \left(\frac{r}{r_\Sigma}\right)^3 \nonumber \\ 
& - & r^3 \int^{r_\Sigma}_r{\frac{e^{(\nu+\lambda)/2}}{r} \left[\frac{\kappa}{2} 
\left(T^1_1 - T^2_2\right)
- E
 \right] dr} \nonumber \\
& - & r^3 \int^{r_\Sigma}_r{ 
\frac{e^{(\lambda-\nu)/2}}{2r}\left(\ddot\lambda + \frac{\dot\lambda^2}{2} 
- \frac{\dot\lambda \dot\nu}{2}\right) 
  dr},
\label{emte}
\end{eqnarray}
or, using (\ref{Wint})
\begin{eqnarray}
m_T & = & (m_T)_\Sigma \left(\frac{r}{r_\Sigma}\right)^3 \nonumber \\ 
& - & r^3 \int^{r_\Sigma}_r{e^{(\nu+\lambda)/2} \left[\frac{\kappa}{r} 
\left(T^1_1 - T^2_2\right)
+ \frac{1}{r^4} \int^r_0{\frac{\kappa}{2} \tilde{r}^3 (T^0_0)' d\tilde{r}} 
 \right] dr} \nonumber \\
& - & r^3 \int^{r_\Sigma}_r{ 
\frac{e^{(\lambda-\nu)/2}}{2r}\left(\ddot\lambda + \frac{\dot\lambda^2}{2} 
- \frac{\dot\lambda \dot\nu}{2}\right) 
  dr}.
\label{emtebisbis}
\end{eqnarray}

For the three cases considered above (i.e.  the static regime, the quasi--static regime, and  immediately after the system departs from equilibrium) we have  $T^0_0=\tilde \rho$ and $T^2_2 - T^1_1=\Pi$. Thus, in any of these cases,  (\ref{emtebisbis}) expresses the Tolman mass of a sphere of radius $r$ interior to $\Sigma$, in terms of its value in  the case of a homogeneous energy density and locally isotropic  fluid in equilibrium (first term), plus the change induced by density inhomogeneity and local anisotropy (second term), plus changes derived from the fact that the system  is not in equilibrium (last term).  For a discussion on this last term see \cite{inh}.We shall, come back to this expression in Sec.IV.

The important point to stress here is that the second  integral in ({\ref{emte}) (or (\ref{emtebisbis})) describes the contribution of density inhomogeneity and local anisotropy of pressure to the Tolman mass. It is also worth noticing that when the system is evaluated immediately after its  departure 
from equilibrium, the value of $\omega$ remains unchanged. 
Therefore the physical meaning of $m_T$, as the active 
gravitational mass obtained for the static (and quasi-static) 
case, may be safely extrapolated to the non-static case within 
that (hydrostatic) time-scale.

\subsection{Structure and evolution equations}
As shown in \cite{Hetal} the following set of equations may be derived to describe the self--gravitating fluid:

\begin{equation}
\tilde \rho ^\ast+(\tilde \rho
+\tilde P_r)\theta=\frac{2}{3}(\theta+\frac{\sigma}{2})\Pi-\tilde q^\dagger-2\tilde q(a+\frac{s^1}{r}),
\label{a'}
\end{equation}

\begin{equation}
\tilde P^{\dagger}_r+(\tilde \rho+\tilde
P_r)a+\frac{2s^1}{r}\Pi=\frac{\tilde q}{3}(\sigma-4 \theta)-\tilde
q^\ast,
\label{b'}
\end{equation}

\begin{equation}
\theta ^\ast+\frac{\theta^2}{3}+\frac{\sigma^2}{6}-a^\dagger
-a^2-\frac{2as^1}{r}=-\frac{\kappa}{2}(\tilde \rho+3\tilde P_r)+\kappa \Pi,
\label{c'}
\end{equation}

\begin{equation}
(\frac{\sigma}{2}+\theta)^\dagger=-\frac{3\sigma s^1}{2r}+\frac{3\kappa}{2}
\tilde q,
\label{d'}
\end{equation}

\begin{equation}
E+\frac{\kappa}{2} \Pi =-a^\dagger -a^2-\frac{\sigma ^\ast}{2}-\frac{\theta
\sigma}{3}+\frac{as^1}{r}+\frac{\sigma^2}{12},
\label{e'}
\end{equation}

\begin{equation}
(\frac{\kappa}{2} \tilde P_r+\frac{3m}{r^3})(\theta +\frac{\sigma}{2})+(E
-\frac{\kappa}{2}\Pi +\frac{\kappa}{2} \tilde \rho)^\ast=-\frac {3\kappa s^1}{2r}\tilde q,
\label{f'}
\end{equation}

\begin{equation}
(E+\frac{\kappa}{2}\tilde \rho-\frac{\kappa}{2}
\Pi)^\dagger=\frac{3s^1}{r}(\frac{\kappa}{2}\Pi-E)+\frac{\kappa}{2} \tilde
q(\frac{\sigma}{2}+\theta),
\label{g'}
\end{equation}

\begin{equation}
\frac{3m}{r^3}=\frac{\kappa}{2} \tilde \rho +\frac{\kappa}{2}(P_\bot -\tilde P_r)+E,
\label{h'}                   
\end{equation}
with $f^\dagger=f_{,\alpha}s^\alpha$,
$f^\ast=f_{,\alpha}u^\alpha$, $a^{\alpha}=as^{\alpha}$, and 
\begin{equation}
\frac{\sigma}{2}+\theta=\frac{3\omega s^1}{r}.
\label{sigmateta}
\end{equation}

These equations are not 
independent  and, of course, provide no more information than the one  contained in Einstein equations,  however  we present them all, since depending on the problem under consideration, it may
be more advantageous using one set instead of the other. 

The first two equations comes from the  ``conservation'' equations $T^\mu_{\nu;\mu}=0$. Equations (\ref{c'}) (Raychaudhuri equation) and (\ref{d'}) are derived from the Ricci identitities, whereas  equation (\ref{e'}) is a consequence of  (\ref{Efeq}), (\ref{34}) and Ricci identities. The next two equations follow from  the Bianchi identities writen in terms of the Weyl tensor (see \cite{Hetal} for details). Finally, (\ref{h'}) is just  (\ref{66}).

We shall next present the orthogonal splitting of the Riemann tensor, and express it in terms of the variables considered so far.

\section{THE ORTHOGONAL SPLITTING OF THE RIEMANN TENSOR}
The orthogonal splitting of the Riemann tensor was first considered by Bel \cite{bel1}, here we shall follow closely  (with some changes) the notation  in  \cite{parrado}.

Thus following Bel, let us introduce the following tensors:
\begin{equation}
Y_{\alpha \beta}=R_{\alpha \gamma \beta \delta}u^{\gamma}u^{\delta},
\label{electric}
\end{equation}
\begin{equation}
Z_{\alpha \beta}=^{*}R_{\alpha \gamma \beta
\delta}u^{\gamma}u^{\delta}= \frac{1}{2}\eta_{\alpha \gamma
\epsilon \rho} R^{\epsilon \rho}_{\quad \beta \delta} u^{\gamma}
u^{\delta}, \label{magnetic}
\end{equation}
\begin{equation}
X_{\alpha \beta}=^{*}R^{*}_{\alpha \gamma \beta \delta}u^{\gamma}u^{\delta}=
\frac{1}{2}\eta_{\alpha \gamma}^{\quad \epsilon \rho} R^{*}_{\epsilon
\rho \beta \delta} u^{\gamma}
u^{\delta},
\label{magneticbis}
\end{equation}
with
$R^{*}_{\alpha \beta \gamma \delta}=\frac{1}{2}\eta_{\epsilon \rho \gamma \delta}R_{\alpha \beta}^{\quad \epsilon \rho}$.

It can be shown that the Riemann tensor  can be expressed through  these tensors in what is called the orthogonal splitting of the Riemann tensor (see \cite{parrado} for details).
Now, instead of using the explicit form of the splitting of Riemann tensor (eq.(4.6) in \cite{parrado}), we shall proceed as follows. 

Using the Einstein equations we may write (\ref{34}) as 
\begin{equation}
R^{\alpha \gamma}_{\quad \beta\delta}=C^{\alpha\gamma}_{\quad
\beta \delta}+2\kappa T^{[\alpha}_{\,\,
[\beta}\delta^{\gamma]}_{\,\, \delta]}+\kappa T(\frac{1}{3} \delta
^{\alpha}_{\,\, [\beta}\delta^{\gamma}_{\,\, \delta]}-\delta
^{[\alpha}_{\quad [\beta}\delta^{\gamma]}_{\,\,
\delta]}),\label{RiemannT}
\end{equation}
then feeding back (\ref{24'}) into  (\ref{RiemannT}) we split the Riemann tensor as 
\begin{equation}
R^{\alpha \gamma}_{\,\, \beta \delta}=R^{\alpha
\gamma}_{(I)\,\,\beta\delta}+R^{\alpha
\gamma}_{(II)\,\,\beta\delta}+R^{\alpha\gamma}_{(III)\,\,\beta\delta},
\label{Riemann}
\end{equation}
where
\begin{equation}
\left. \begin{array}{l}
 R^{\alpha \gamma}_{(I)\,\,\beta \delta}=2\kappa \tilde \rho
u^{[\alpha}u_{[\beta}\delta^{\gamma]}_{\,\,\delta]}-2\kappa \hat
Ph^{[\alpha}_{\,\,[\beta}\delta^{\gamma]}_{\,\, \delta]}+\kappa
(\tilde \rho-3\hat
P)(\frac{1}{3}\delta^{\alpha}_{\,\,[\beta}\delta^{\gamma}_{\,\,\delta]}
-\delta^{[\alpha}_{\,\,[\beta}\delta^{\gamma]}_{\,\,\delta]})
\\
\\
R^{\alpha \gamma}_{(II)\,\,\beta\delta}=2\kappa (\Pi^{[\alpha}_{\,\,
[\beta}\delta^{\gamma]}_{\,\, \delta]}+ \tilde qs^{[\alpha}u_{\,\,
[\beta}\delta^{\gamma]}_{\,\, \delta]}+\tilde qu^{[\alpha}s_{\,\,
[\beta}\delta^{\gamma]}_{\,\, \delta]})
\\
\\
R^{\alpha
\gamma}_{(III)\,\,\beta\delta}=4u^{[\alpha}u_{[\beta}E^{\gamma]}_{\,\,\,
\delta]}-\epsilon^{\alpha \gamma}_{\quad
\mu}\epsilon_{\beta\delta\nu}E^{\mu\nu}
\label{RiemannH}
\end{array} \right \}
\end{equation}
with
\begin{eqnarray}
\epsilon_{\alpha\gamma\beta}=u^\mu\eta_{\mu\alpha\gamma\beta},\quad
\epsilon_{\alpha\gamma\beta}u^\beta=0 \label{conn},
 \end{eqnarray} 
and where the vanishing, due to the spherical symmetry, of the  magnetic part of the Weyl tensor ($H_{\alpha \beta}=^{*}C_{\alpha \gamma \beta
\delta}u^{\gamma}u^{\delta}$)  has been used.

From (\ref{conn}) it follows that 
$\epsilon^{\mu \gamma \nu}\epsilon_{\nu \alpha \beta}=u^\sigma u^\rho \eta_{\rho} ^{ \mu \gamma \nu}\eta_{\sigma \nu \alpha \beta}$, producing
\begin{equation}
\epsilon^{\mu \gamma \nu}\epsilon_{\nu \alpha \beta}
=\delta^\gamma_\alpha h^\mu_\beta-\delta^\mu_\alpha h^\gamma_\beta+u_\alpha(u^\mu \delta^\gamma_\beta-\delta^\mu_\beta u^\gamma),
\label{neps}
\end{equation}
or, contracting  $\alpha$ with $\mu$ in (\ref{neps}) 
\begin{equation}
\epsilon^{\mu \gamma \nu}\epsilon_{\nu \mu \beta}
=-2 h^\gamma_\beta.
\label{nepsII}
\end{equation}

Using the results above, we can now  find the explicit expressions for the three tensors $Y_{\alpha \beta}, Z_{\alpha \beta}$ and  $X_{\alpha \beta}$ in terms of  the physical variables, we obtain

\begin{equation}
Y_{\alpha\beta}=\frac{\kappa}{6}(\tilde\rho+3\hat
P)h_{\alpha\beta}+\frac{\kappa}{2}\Pi_{\alpha\beta}+E_{\alpha\beta},\label{Y}
\end{equation}

\begin{equation}
Z_{\alpha\beta}=\frac{\kappa}{2}\tilde
qs^{\mu}\epsilon_{\alpha\mu\beta},\label{Z}
\end{equation}

and

\begin{equation}
X_{\alpha\beta}=\frac{\kappa}{3}\tilde \rho
h_{\alpha\beta}+\frac{\kappa}{2}
\Pi_{\alpha\beta}-E_{\alpha\beta}.\label{X}
\end{equation}
From the above, we can obtain expressions for two quantities endowed with  a profound physical meaning (see \cite{parrado} , \cite{seno}, \cite{h} and references therein). They appear in the orthogonal splitting of the Bel tensor \cite{bel2}. One of them is the Bel superenergy, defined by 
\begin{equation}
\bar {W}=\frac{1}{2}(X^{\alpha \beta}X_{\alpha \beta}+Y^{\alpha \beta}Y_{\alpha \beta})+Z^{\alpha \beta}Z_{\alpha \beta},
\label{s1}
\end{equation}
the other is the super--Poynting vector, defined as 
\begin{equation}
\bar{P}_{\alpha}=\epsilon_{\alpha \beta \gamma}(Y_{\delta}^{\gamma}Z^{\beta \delta}-X^{\gamma}_{\delta}Z^{\delta \beta}).
\label{s6}
\end{equation}

Similar quantities may also be defined from  the orthogonal splitting of the Bel--Robinson tensor \cite{bel3}. However it should be noticed that due to the vanishing of the magnetic part of the Weyl tensor in the spherically symmetric case, the super--Poynting vector associated to the Bel--Robinson tensor vanishes. Also, in the conformally flat case the superenergy associated to the  Bel--Robinson tensor  vanishes. In other words, these two definitions cover a much wider range of situations, when defined from the Bel tensor. Of course, in vacuum both set of definitions coincide.
Then, from (\ref{s1}) and (\ref{s6}), using (\ref{Y}, \ref{Z},  
\ref{X}) we find 
\begin{equation}
\bar W=\frac{\kappa^2}{24}(5\tilde \rho^2+6\tilde \rho \hat
P+9\hat P^2+4\Pi^2)+\frac{2}{3}E^2 +\frac{\kappa^2}{2}\tilde
q^2,\label{w1}
\end{equation}

\begin{equation}
\bar P_{\alpha}=\frac{\kappa ^2}{2}\tilde q(\tilde \rho+\tilde
P_r)s_\alpha. \label{Poynting1}
\end{equation}

Observe that the superenergy associated to the Bel--Robinson tensor ($W$), which is defined by 
\begin{equation}
W=E^{\alpha \beta}E_{\alpha \beta},\label{s1b}
\end{equation}
(assuming the magnetic part of the Weyl tensor vanishes, as it happens in our case), takes the form
\begin{equation}
W=\frac{2}{3}E^2.
\label{52bis}
\end{equation}
\noindent Combining  (\ref{52bis}) with  (\ref{w1}) we found

\begin{equation}
\bar W-W=\frac{\kappa^2}{24}(5\tilde \rho^2+6\tilde \rho \hat
P+9\hat P^2+4\Pi^2) +\frac{\kappa^2}{2}\tilde q^2.\label{ww}
\end{equation}
It is also worth noticing that the super--Poynting vector vanishes if and only if there is not dissipative flux. This fact  clearly illustrates the physical meaning of this vector, and  fully justifies its name.

\subsection{Five relevant scalars}
 
We shall now derive five scalars quantities (hereafter referred to as structure scalars), in terms of which we shall write our equations (\ref{a'})--(\ref{h'}).

Let us first observe that  tensors  $X_{\alpha \beta}$ and $Y_{\alpha \beta}$ can be splitted in terms of their traces and  the corresponding trace--free tensor, i.e.
\begin{equation}
X_{\alpha \beta}=\frac{1}{3}Tr X h_{\alpha \beta}+ X_{<\alpha \beta>},
\label{esn}
\end{equation}
with $Tr X=X^\alpha_\alpha$ and,
\begin{equation}
X_{<\alpha \beta>}=h^\mu_\alpha h^\nu_\beta(X_{\mu \nu}-\frac{1}{3}Tr X h_{\mu \nu}).
\label{esnII}
\end{equation}
From (\ref{X}) we have
\begin{equation}
TrX\equiv X_T=\kappa \tilde \rho,
\label{esnIII}
\end{equation}
and
\begin{equation}
X_{<\alpha \beta>}=X_{TF}(s_\alpha s_\beta+\frac{h_{\alpha \beta}}{3}),
\label{esnIV}
\end{equation}
where  
\begin{equation}
X_{TF}\equiv (\frac{\kappa \Pi}{2}-E).
\label{esnIVn}
\end{equation}

In a similar way we obtain 
\begin{equation}
TrY\equiv Y_T=\frac{\kappa}{2}(\tilde \rho+3\tilde P_r-2\Pi),
\label{esnV}
\end{equation}
and
\begin{equation}
Y_{<\alpha \beta>}=Y_{TF}(s_\alpha s_\beta+\frac{h_{\alpha \beta}}{3}),
\label{esnVI}
\end{equation}
with  
\begin{equation}
Y_{TF}\equiv (\frac{\kappa \Pi}{2}+E).
\label{defYTF}
\end{equation}
Finally a fifth scalar may be defined  from (\ref{Z}) as 
\begin{equation}
Z=
\sqrt{Z_{\alpha\beta}Z^{\alpha\beta}}=\frac{\kappa}{\sqrt{2}}\tilde q.
\label{zz}
\end{equation}
From the above it follows that local anisotropy of pressure is  determined by  $X_{TF}$ and $Y_{TF}$
by
\begin{equation}
\kappa \Pi=X_{TF} + Y_{TF}.
\label{defanisxy}
\end{equation}
We can now rewrite equations (\ref{a'})-(\ref{h'}) in terms of our five structure scalars ($X_T, X_{TF}, Y_T, Y_{TF}, Z$):
\begin{equation}
\frac{\kappa}{2}\tilde \rho
^\ast+\frac{1}{3}(X_T+Y_T+X_{TF}+Y_{TF})\theta=\frac{1}{3}(\theta+\frac{\sigma}{2})(X_{TF}+Y_{TF})-\sqrt{2}
( \frac{Z^\dagger}{2} +aZ+\frac{s^1}{r}Z), \label{a''}
\end{equation}
\begin{equation}
\frac{\kappa}{2} \tilde P_r ^\dagger
+\frac{1}{3}(X_T+Y_T+X_{TF}+Y_{TF})a+\frac{s^1}{r}(X_{TF}+Y_{TF})=\frac{\sqrt{2}}{3}(\frac{\sigma}{2}-2\theta)Z-\sqrt{2}\frac{Z^\ast}{2},\label{b''}
\end{equation}
\begin{equation}
\theta ^\ast+\frac{1}{3}\theta^2+\frac{\sigma
^2}{6}-a^\dagger-a^2-\frac{2}{r}as^1=-Y_T,\label{c''}
\end{equation}
\begin{equation}
(\frac{\sigma}{2}+\theta)^\dagger=-\frac{3\sigma s^1}{2r}+\frac{3\sqrt{2}}{2}Z\label{d''n}
\end{equation}

\begin{equation}
a^\dagger+a^2+\frac{\sigma^\ast}{2}+\frac{1}{3}\theta
\sigma-\frac{a}{r}s^1-\frac{1}{12}\sigma^2=-Y_{TF},\label{d''}
\end{equation}

\begin{equation}
\frac{1}{3}\left[(Y_T+Y_{TF})-2X_{TF}+X_T\right](\theta+
\frac{1}{2}\sigma)+(\frac{X_T}{2}-X_{TF})^\ast=-\frac{3}{2r}s^1\sqrt{2}Z, \label{e''}
\end{equation}

\begin{equation}
(\frac{\kappa}{2}\tilde
\rho-X_{TF})^\dagger=\frac{3s^1}{r}X_{TF}+\frac{\sqrt{2}}{2}Z(\frac{\sigma}{2}+\theta), \label{f''}
\end{equation}

\begin{equation}
\frac{3m}{r^3}=\frac{X_T}{2}-X_{TF}. \label{h''}
\end{equation}

Whereas for the Bel superenergy and the super--Poynting vector  we find

\begin{equation}
\bar
W=\frac{1}{6}(X_T^2+Y_T^2)+\frac{1}{3}(X_{TF}^2+Y_{TF}^2)+Z^2, \label{wxy}
\end{equation}
 and
 \begin{equation}
\bar P_{\alpha}=\frac{\sqrt{2}}{3}Z(X_T+Y_T+X_{TF}+Y_{TF})s_\alpha. \label{Poynting1bis}
\end{equation}
\subsection{On the physical meaning of the structure scalars}
Let us now focus on the physical meaning of the scalars introduced in the previous subsection.

The physical meaning of  $X_T$  and $Z$ is evident and does not require further clarification. 

Let us now consider  $X_{TF}$. From  (\ref{f''}) it follows  that in the absence of dissipation  ($Z=0$), (using the regular center condition),
\begin{equation}
 \tilde\rho^{\dagger}=0\Leftrightarrow X_{TF}=0.
 \label{arrown}
 \end{equation}

 In other words, in the absence of dissipation,   $X_{TF}$ controls inhomogeneities in the energy density. 

The role of density inhomogeneities in the collapse of dust \cite{MeTa}
and in particular in the formation of naked singularities \cite{VarI}--\cite{VarVIII},
has been extensively discussed in the literature.

In the non--dissipative locally isotropic case, we obtain from (\ref{f''}), $\tilde\rho^{\dagger}=0\Leftrightarrow E=0$. This link between the Weyl tensor and energy density inhomogeneity  and the fact that  tidal forces tend to make the gravitating
fluid more inhomogeneous as the evolution proceeds, led Penrose to propose    a gravitational arrow of time in terms of the Weyl tensor \cite{Pe} (see also \cite{Wa} and references therein).

 However, the fact that such a relationship is no longer valid in the presence
of local anisotropy of the pressure and/or
dissipative processes, already discussed in \cite{Hetal}, explains its
failure in scenarios where the above-mentioned  factors are present \cite{arrow}.

Here we see that the single scalar $X_{TF}$ (in the absence of dissipation)  controls density inhomogeneities, and therefore should be the fundamental ingredient in the definition of a gravitational arrow of time. If dissipative processes are present, the scalar $Z$ should be incorporated into that definition.

To establish the physical meaning of  $Y_T$  and $Y_{TF}$ let us get back to equations (\ref{emte}), using (\ref{defYTF}) (for the three cases considered in 2.3.1) we get

\begin{eqnarray}
m_T & = & (m_T)_\Sigma \left(\frac{r}{r_\Sigma}\right)^3 \nonumber \\
& + & r^3 \int^{r_\Sigma}_r{\frac{e^{(\nu+\lambda)/2}}{r} Y_{TF}dr} \nonumber \\
& - & r^3 \int^{r_\Sigma}_r{
\frac{e^{(\lambda-\nu)/2}}{2r}\ddot\lambda  dr}.
\label{emtebis}
\end{eqnarray}

 Comparing the above expression with (\ref{emtebisbis}) we see that $Y_{TF}$ describes the influence of the local anisotropy of pressure and density inhomogeneity on the Tolman mass. It is also worth recalling that $Y_{TF}$,  together with $X_{TF}$, determines the local anisotropy of the fluid distribution.

Finally, observe that for a system in equilibrium or quasi--equilibrium, the Tolman mass (\ref{Tolin}) becomes
 \begin{eqnarray}
m_T = & &  \frac{\kappa}{2} \int^{r}_{0}{r^2 e^{(\nu+\lambda)/2}
\left(T^0_0 - T^1_1 - 2 T^2_2\right) dr},
\label{TolinII}
\end{eqnarray}
which for those two regimes may be written as
\begin{eqnarray}
m_T = & &   \int^{r}_{0}{r^2 e^{(\nu+\lambda)/2}
Y_Tdr}.
\label{TolinIII}
\end{eqnarray}

Thus  $Y_T$ appears to be proportional to the  Tolman mass ``density'' for systems in equilibrium or quasi--equilibrium.
\section{ALL STATIC ANISOTROPIC SPHERES}
In this section we shall restrict ourselves to static sytems. We shall  see how the metric corresponding to any static anisotropic sphere can be expressed in terms of the structure scalars.
 We shall explore three possible alternatives.
\subsection{First alternative}
From (\ref{rieman}) and (\ref{h''}) it follows at once that 

\begin{equation}
e^{-\lambda}=1-\frac{2}{3}r^2(\frac{1}{2}X_T-X_{TF}).\label{lambda'}
\end{equation}
Next, using (\ref{c''}) and (\ref{d''}) in the static case, we may write
\begin{equation}
a=\frac{r}{3s^1}(Y_{TF}+Y_{T}).\label{ac''}
\end{equation}
\noindent On the other hand we have in the static case 
\begin{equation}
a=e^{-\frac{\lambda}{2}}\frac{\nu ^\prime}{2}, \qquad
s^1=e^{-\frac{\lambda}{2}}\label{sa},
\end{equation}
where (\ref{15a}) and (\ref{ese}) have been used. Feeding back  (\ref{sa})  into  (\ref{ac''}), and integrating we obtain
\begin{equation}
e^{\nu}=Ce^{\int\frac{2r}{3}\frac{Y_{TF}+Y_T}{1-\frac{2r^2}{3}(\frac{1}{2}X_T-X_{TF})}dr},
\label{nu'}
\end{equation}
 where $C$ is a constant of integration easily obtained from (\ref{enusigma}).
 \noindent 
 
 Then, using  (\ref{lambda'}) and (\ref{nu'}), the line element 
  (\ref{metric}) in the static case may be writen as  \begin{equation}
ds^2=Ce^{\int\frac{2r}{3}\frac{Y_{TF}+Y_T}{1-\frac{2r^2}{3}(\frac{1}{2}X_T-X_{TF})}dr}dt^2-\frac{1}{1-\frac{2}{3}r^2(\frac{1}{2}X_T-X_{TF})}dr^2
-r^2(d\theta ^2+\sin^2 \theta d\phi^2)\label{metric2f}.
\end{equation}

Thus we see that all possible  spacetimes generated by  static anisotropic fluids, are fully determined  by two scalars, namely: $Y_{TF}+Y_T$  and $\frac{1}{2}X_T-X_{TF}$.
\subsection{Second alternative}
Alternatively we may proceed as follows. 
From (\ref{m}), (\ref{esnIII}) and (\ref{h''}) we obtain
\begin{equation}
m(r)=\frac{r^3}{3}\left(\frac{m^\prime}{r^2}-X_{TF}\right),\label{mstat''}
\end{equation}
\noindent which after integration produces \begin{equation}
m(r)=r^3\left(\int \frac{X_{TF}}{r} dr +c_1\right).\label{mXTF}
\end{equation}
Or, using (\ref{rieman})

\begin{equation}
e^{-\lambda}=1-2r^2\left(\int \frac{X_{TF}}{r} dr+c_1\right),\label{lambda'}
\end{equation}
where the constant of integration $c_1$ may be easily calculated from (\ref{elambdasigma}).

Next, the field equation (\ref{2}) in the  static case may be writen as 
\begin{equation}
\frac{\kappa}{2}P_r=\frac{1}{2}\frac{e^{-\lambda}-1}{r^2}+e^{-\lambda}\frac{\nu^\prime}{2r},\label{3lja}
\end{equation}
\noindent or, using (\ref{rieman}), (\ref{esnV}), (\ref{defYTF}),  (\ref{ac''}) and (\ref{sa})
\begin{equation}
\frac{\kappa}{2}P_r+\frac{m}{r^3}=e^{-\lambda}\frac{\nu^\prime}{2r}=Y_h,\label{nuYh}
\end{equation}
 with
\begin{equation}
Y_h=\frac{1}{3}(Y_T+Y_{TF}).
\label{defYh}
\end{equation}

We can now integrate  (\ref{nuYh}) to obtain
\begin{equation}
e^{\nu}=c_2e^{\int \frac{2r Y_h}{1-2r^2(\int \frac{X_{TF}}{r} dr +c_1)}dr},
\label{nuYh'}
\end{equation}
where the constant of integration $c_2$  may be obtained from (\ref{enusigma}).

 \noindent Thus using (\ref{lambda'}) and (\ref{nuYh'}),
 the line element  (\ref{metric}) in the static case may be writen as \begin{equation}
ds^2=c_2e^{\int \frac{2r Y_h}{1-2r^2(\int \frac{X_{TF}}{r} dr
+c_1)}dr}dt^2-\frac{1}{1-2r^2(\int \frac{X_{TF}}{r} dr+c_1)}dr^2 -r^2(d\theta
^2+\sin^2 \theta d\phi^2),\label{metric2}
\end{equation}
allowing the representation of   all possible metrics in terms of two scalar functions, $Y_h$ and $X_{TF}$.
\subsection{Third alternative}
In  the two previous alternatives we have seen that all line elements corresponding to an anisotropic fluid may be determined by two scalars. However  in both cases these two scalars are a combination of  four (alternative I) or three (alternative II)  of our structure scalars. Here  we shall present a third alternative, which allows for describing any line element in terms of only two structure scalars, namely  $X_{TF}$ and $Y_{TF}$.

 Now, in the static case we may write from (\ref{defE}),
\begin{equation}
E=-\frac{e^{-\lambda}}{2}\left[\frac{\nu ^{\prime
\prime}}{2}+(\frac{\nu ^\prime}{2})^2+\frac{\nu
^\prime}{2}(-\frac{\lambda ^\prime}{2}-\frac{1}{r})+\frac{\lambda
^\prime}{2r}+\frac{1-e^{\lambda}}{r^2}\right],\label{E}
\end{equation}
then introducing new variables
\begin{equation}
y=e^{-\lambda};\qquad \frac{\nu ^\prime}{2}=\frac{ u^\prime}{u},
\label{nv}
\end{equation}
(\ref{E}) becomes
\begin{equation}
y^\prime +2y \frac{u^{\prime
\prime}-\frac{u^\prime}{r}+\frac{u}{r^2}}{u^\prime-\frac{u}{r}}=\frac{2u(1-2r^2E)}{r^2(u^\prime-\frac{u}{r})},\label{E1}
\end{equation}
which after integration yields
\begin{equation}
y=e^{-\int k(r)dr}(\int e^{\int k(r) dr} f(r) dr+C_1),\label{E2}
\end{equation}
with
$$k(r)=2 \frac{d}{dr}\left[ln(u^\prime-\frac{u}{r})\right], \quad  \quad
f(r)=\frac{2u(1-2r^2E)}{r^2(u^\prime-\frac{u}{r})},$$
and where $C_1$ is a constant of integration easily obtained from the junction conditions.

Then, getting back to original  variables, (\ref{E2}) becomes
\begin{equation}
\frac{\nu^\prime}{2}-\frac{1}{r}=\frac{e^{\frac{\lambda}{2}}}{r}\sqrt{(1-2r^2E)+e^{-\nu}r^2C_1+e^{-\nu}
r^2\int (2r^2E)^\prime \frac{e^\nu}{r^2}dr}.\label{E3}
\end{equation}
\noindent  Next let us introduce the new variable $z$ by, 
\begin{equation}
e^{\nu}=\frac{e^{2\int zdr}}{r^2},
\label{zintr}
\end{equation}
producing
\begin{equation}
z(r)=\frac{\nu^\prime}{2}+\frac{1}{r}\label{zw}.
\end{equation}
Using (\ref{zintr}) and (\ref{zw}) in (\ref{E3}) we get a  a link between $E$  and $z$
 \begin{equation}
z(r)=\frac{2}{r}+\frac{e^{\frac{\lambda}{2}}}{r}\sqrt{(1-2r^2E)+r^4e^{-\int
2 z(r)dr}C_1+r^4e^{-\int 2z(r)dr} \int (2r^2E)^\prime\frac{
e^{\int 2z(r)dr}}{r^4}dr}.\label{E4}
\end{equation}
\noindent Next, from field equations (\ref{2}) and ({\ref{3'}) it follows
\begin{equation}
\kappa \Pi=e^{-\lambda}\left[-\frac{\nu^{\prime
\prime}}{2}-(\frac{\nu^{\prime}}{2})^2+\frac{\nu^{\prime}}{2r}+\frac{1}{r^2}\right]+e^{-\lambda}\frac{\lambda^{\prime}}{2}(
\frac{\nu^{\prime}}{2}+\frac{1}{r})-\frac{1}{r^2},\label{anisnuevo}
\end{equation}

\noindent which in terms of the variables $z$ and $y$ introduced above, becomes
\begin{equation}
y^{\prime}+y(\frac{2z^\prime}{z}+2z-\frac{6}{r}+\frac{4}{r^2
z})=-\frac{2}{z}(\frac{1}{r^2}+\kappa \Pi). \label{eq1nuevo}
\end{equation}

\noindent Integrating (\ref{eq1nuevo}) we obtain for  $\lambda$:

\begin{equation}
e^{\lambda (r)}=\frac{z^2(r) e^{\int(\frac{4}{r^2 z(r)}+2z(r))dr}}
{r^6(-2\int\frac{z(r)(1+\kappa \Pi (r)r^2 ) e^{\int(\frac{4}{r^2
z(r)}+2z(r))dr}}{r^8}dr+C)},\label{lambdaIIIp}
\end{equation}
where $C$ is a constant of integration.

In terms of $X_{TF}$ and  $Y_{TF}$, equations (\ref{E4}) and (\ref{lambdaIIIp}) may be writen as:
\begin{equation}
z(r)=\frac{2}{r}+\frac{e^{\frac{\lambda}{2}}}{r}\sqrt{\left[1-r^2(Y_{TF}-X_{TF})\right]+r^4e^{-\int
2 z(r)dr}(C_1+ \int [r^2(Y_{TF}-X_{TF})]^\prime\frac{
e^{\int 2z(r)dr}}{r^4}dr},\label{E4bis}
\end{equation}
and
\begin{equation}
e^{\lambda (r)}=\frac{z^2(r) e^{\int(\frac{4}{r^2 z(r)}+2z(r))dr}}
{r^6(-2\int\frac{z(r)\left[1+(X_{TF}+Y_{TF})r^2 \right] e^{\int(\frac{4}{r^2
z(r)}+2z(r))dr}}{r^8}dr+C)}.\label{lambdaIII}
\end{equation}
\noindent Thus,  given  $X_{TF}$ and $Y_{TF}$ and using (\ref{lambdaIII}) in (\ref{E4bis}) we obtain $z$, which by virtue of (\ref{zintr}) allows to find $\nu$. Then using $z$ in (\ref{lambdaIII}) determines $\lambda$.

This approach is essentially  equivalent to the method for obtaining static anisotropic solutions presented in \cite{a37} (see also \cite{lakebis}).

As a simple example let us consider all conformally flat anisotropic fluids. Conformal flatness implies $Y_{TF}=X_{TF}$. Feeding back this condition into (\ref{E4bis}) we obtain for $z$
\begin{equation}
z=\frac{2}{r}+\frac{e^{\frac{\lambda}{2}}}{r}\tanh(\int
\frac{e^{\frac{\lambda}{2}}}{r}dr).\label{zlambda}
\end{equation}

In order to specify  a single solution we have to provide another condition on our scalars. Thus for example, if we assume further the energy density to be constant, then $X_{TF}=0$ implying $Y_{TF}=0$, leading to the well known Schwarzschild interior solution.
\section{CONCLUSIONS}

We have presented a systematic study of  spherically symmetric self--gravitating  relativistic fluids, based on  scalars functions derived from the orthogonal splitting of the Riemann tensor. In the most general case (dissipative and anisotropic fluid) we have five scalars, which reduce to two, in the case of dissipationless dust and static anisotropic fluids, and to one  for  static isotropic fluids.

The motivation to present such a study and to consider further these scalars in the study of self--gravitating objects stems from  their distinct physical meaning. As we have seen, two of them ($X_T$ and $Z$) define the energy  density and the dissipative flux, respectively. In the absence of dissipation, $X_{TF}$ controls the inhomogeneity of energy density and therefore is the relevant quantity in any definition of a gravitational arrow of time {\it a la} Penrose. Of course, in the dissipative case $Z$ must also enter into that definition.

The two scalars $Y_{TF}$ and $Y_T$ are related in a conspicuous way to the Tolman mass. On the one hand $Y_{TF}$ describes the influence of, both, energy density inhomogeneity and local anisotropy of pressure, on the Tolman mass. On the other hand $Y_T$ acts as a Tolman mass density. It is  worth noticing that these two scalars are the only ones that appear in the ``kinematical'' equations (\ref{c''}) and (\ref{d''}). Also observe that $Z$ is the only structure scalar appearing in (\ref{d''n}).

In the static case, Einstein equations reduce to three ordinary differential equations for five variables ($\rho, P_r, P_\bot, \nu, \lambda $), implying that any specific solution is determined by two scalar functions. This was illustrated in the three subsections above. Furthermore, these two scalars functions may be $X_{TF}$ and $Y_{TF}$ as  shown in the third alternative developped in subsection 4.3. This  brings out further the physical relevance of the structure scalars.

\section{Acknowledgements}
J.O  acknowledge financial assistance under grant
BFM2003-02121 (M.C.T. Spain).

\end{document}